          [2004/06/22 1.2 (KOF); 2001/04/25 1.1 (PWD)]

\documentclass[a4paper,twocolumn]{Gaia2004} 
\usepackage{times}      
\usepackage{amssymb}
\usepackage{epsfig}     
\usepackage{natbib}     
\title{Variability Analysis: Detection and Classification}

\author{Laurent Eyer}
\affil{Observatoire de Gen\`eve, CH-1290 Sauverny}

\bibpunct{(}{)}{;}{a}{}{,}  

\begin{document}


\maketitle

\begin{abstract}
  
  Gaia mission will offer an exceptional opportunity to perform variability studies. The data
  homogeneity, its optimised photometric systems, composed of 11 medium and 4-5 broad bands,
  the high photometric precision in G band of one milli-mag for $V=13-15$, the radial velocity
  measurements and the exquisite astrometric precision for one billion stars will permit a
  detailed description of variable objects like stars, quasars and asteroids.
  
  However the time sampling and the total number of measurements change from one object to
  an other because of the satellite scanning law. The data analysis is a challenge because of
  the huge amount of data, the complexity of the observed objects and the peculiarities of the
  satellite, and needs a thorough preparation.
  
  Experience can be gained by the study of past and present survey analysis and results
  and Gaia should be put in perspective with the future large scale surveys, like Pan-STARRS
  or LSST.

  We present the activities of the Variable Star Working Group and a general plan
  to digest this unprecedented data set, focusing here on the photometry.


\end{abstract}

\section{Introduction}

The logarithmic representation of the Universe by \citet{RGMJ03}\\
(http://www.astro.princeton.edu/$\sim$mjuric/universe/) clearly
shows the breadth and diversity of the Gaia mission Science. 

Gaia will measure targets from the solar system small bodies to the distant quasars, the bulk of objects being 1 billion stars in our Galaxy.

A huge number of the objects will be variable. Indeed asteroids are generally
variable at the level $\delta V~\sim$~0.1-0.7 mag.  Though the QSOs are sometimes
less variable than usually thought \citep{PS05}, a very large fraction of them will
be detected as variable with Gaia.

On the side of stars, there is a Zoo of variables: If we look at the GCVS we may count
more that 100 types and subtypes. It is customary to divide the variable stars in five
main classes:
Pulsating stars, eclipsing binaries, variability induced by rotation, eruptive stars and cataclysmic variable stars. The time scale and amplitude of the variation may be very
different: from short event of seconds to secular evolution and from milli-magnitudes to several magnitudes. The variability may be periodic, or irregular.

The several instrumental devices on board, photometric, astrometric and radial velocity,
will allow a deep investigation on variability induced phenomena. Gaia will allow a global
and precise description of stellar variability in terms of physical quantities (temperature,
gravity, metallicity). On classical topics, Gaia will give access to absolute masses and sizes
for about $10\,000$ eclipsing binaries as pointed out by \citet{TZ03}, the
period-luminostiy-metallicity realtions of Cepheids will greatly benefit from the parallaxes
(cf. Ngeow \& Kanbur, these proceedings). Asteroseismology will also benefit from accurate
parallaxes, because luminosity and radius are stringent test for the stellar structure,
cf. \citet{FF99}.

The G-band photometric and astrometric informations are deduced from the same CCD counts, therefore an interdependence between the two processings exists. As observed in other surveys,
variability may affect astrometric measurements \citep[cf. ][]{DPetal03} and astrometry
(proper motions) may affect photometry by the reduction methods, cf. \citet{LEPW01} or
\citet{CAetal01}. Non uniform light distribution in Long Period Variables (LPVs) may affect
the measured astrometry through photocenter wobbling (Jorissen, private communication).

For these reasons, the variability analysis is an unavoidable step of the Gaia mission to
improve both scientific returns and astrometric quality.

\section{Past Surveys}

In recent years, only the {\sc hipparcos} satellite has done a multi-epoch photometric
all-sky survey with an associated analysis aimed at systematically detecting variability.
The {\sc hipparcos} programme consists of 118\,204 stars among them a survey of 52\,045 stars,
complete down to a well defined limiting magnitude function of the star colour and its
ecliptic latitude. Results on variable stars were published in the volumes 11 and 12 and
among the flags of the main catalogue \citep{ESA97}. The rate of variable stars is very
high, about 10\% with a similar rate in the survey and in the preselected star sample.

Microlensing surveys like EROS, MACHO, OGLE detected and characterised a large number of
variable stars with a lesser detection rate of about 1\%.
To mention one, OGLE-II observed about 40 million stars (\citet{AUetal98}, \citet{AUetal00}, \citet{AUetal02}) and \citet{KZ01} \citet{PWetal02} extracted about 290\,000 variable stars, data
publicly available.

From OGLE and {\sc hipparcos} surveys, we can see that there is a certain time lag between the data acquisition end and the published results of the variability. The OGLE-II survey collected data from 1997 to 2000, published the photometry for the variable objects in 2001 for the
Magellanic clouds and in 2002 for the bulge. In 2004, the database is not fully scientifically exploited yet. {\sc Hipparcos} finished the measurements in 1993, the catalogue was published in 1997, though the full (prepared) variability analysis was completed within a few months.

These examples show that a thorough preparation of the analysis should be made well in advance
in order not to lag with the data flow.

\section{Ongoing and Future Surveys}

The study of variability for Gaia should also be put in perspective with other projects.
A website lists about 90 existing or future robotic telescopes
(http://alpha.uni-sw.gwdg.de/$\sim$hessman/MONET/).

We mention here three ongoing surveys:

-- ASAS \citep[All-Sky Automated Survey, ][]{GP97} is a project which goal is
the photometric monitoring of
approximately 10 million stars brighter than magnitude 14. The photometry is in $V$ and $I$ bands. The data for variable stars is publicly available. There is an alert system.

-- OGLE (Optical Gravitational Lensing Experiment) is continuing to observe the bulge 
  and the Magellanic clouds. It has also dedicated campaigns aiming at
  the detection of planetary transits \citep{AU03a}. OGLE is the first team to have
  detected planetary systems with the transit method. OGLE-III has a real time analysis
  \cite{AU03b}.

-- SDSS (Sloan Digital Sky Survey) provides multi-epoch measurements. In the Northern sky,
the number of epochs is limited to two but thanks to the multi-colour data, it allows
to detect efficiently some variability types cf. \citet{ZI00}, \citet{ZI03}.
We note that in the Southern Survey the number of epochs will be about a dozen.

There are two large scale future surveys which should be mentioned:

-- Pan-STARRS (Panoramic Survey Telescope \& Rapid Response System) will monitor
the sky down to a limiting magnitude of 24 about 20 times per year.
The survey will be executed by four 1.8m telescopes located in Hawaii. Each
telescope will have a 3 deg field of view. The first light of the
first telescope is planned for 2006 and the full system will be
operational in 2008.  The photometric system will have the Sloan g, r,
i bands, a "sloanish" z band and a y band at 1.7 micron. The photometric
precision at the bright end will be about 5 mmag.

-- LSST (Large Synoptic Survey Telescope) will cover down to a limiting magnitude
of 24 in 5 bands 20\,000 square degrees with a single 8.4m telescope.
The first light is planned for 2012.

\section{Photometric Surveys from Space}

Two space missions, COROT and Kepler, will describe as by products
the behaviour of large numbers of small amplitude variable stars thanks to
their high photometric precision. Two remarkable small satellites, WIRE \citep{DB02}
and MOST \citep{GWJM03}, have been exploring the small amplitude variability domain
working both in a single target mode:

-- The CNES/ESA COROT mission (http:// smsc.cnes.fr/COROT/index.htm): The
satellite will be in operation from 2006 to 2009 with a possible
extension. There will be two programs run in parallel, one dedicated
to asteroseismology, the other to planetary transit searches. The exo-planet
program will monitor in total 30\,000-60\,000 stars down to V=15.5 with colour
information, in five different fields, during 150 days each. Measurements
will be transferred to Earth every 8 minutes. After 1 hour of integration, 
mean intensities will have a precision of 700 ppm (parts per million)
at $V=15.5$. The asteroseismology program will monitor, 5
different fields containing one primary target per field brighter
than $V=6$  and a maximum of 9 secondary targets during 150 days.
Finally there will be 4 exploratory programs of 30 days each.

-- The Kepler Mission (NASA, http:// www.Kepler.NASA.gov/): It will be
launched in 2007 and will be operational for 4
years, with a possible 2 year mission extension.  At least 100\,000
preselected main sequence stars, in the magnitude range from 9 to 15,
will be continuously monitored thanks to the Earth-trailing
heliocentric orbit and the position of the observed field selected to
avoid interruptions in the data acquisition. After 6.5 hours of
integration, the mean intensity, for a source of mag V=12, will have a
precision of $<$14 ppm (including instrument, background
and photon shot noises).  Additionally, 200 sources will be monitored
for asteroseismological purposes at a one-minute cadence.  These source will
be changed every 1 to 3 months so as to have monitored all the bright
dwarfs in the FOV for p- and g-modes, about 5\,000 of them.

\section{Gaia Survey}

Gaia will observe 1 billion objects with a mean of $\sim$80 measurements in the Astro-field (G-band
and Broad Band Photometry-BBP), and $\sim$90 measurements in the Medium Band Photometry (MBP)
during 5 years. Possibly a near IR magnitude from the Spectro-field can also be computed.
The photometric systems are still under discussions. The photometric precision in the G-band is
very high from 1 mmag at V$=$13-15 to 0.02 mag at the limiting magnitude V$=$20; for more
details on the photometric performances, see \citet{CJ04}.

\subsection{Gaia Sampling}

The satellite will be operated at the second Lagrangian point. The sampling law is governed by
two main factors:
\begin{enumerate}
  \item The rotation of the satellite on itself in 6 hours.
  \item The precession of the satellite rotation axis on a cone of 50 deg angle to
        the Sun, in 70 days.
\end{enumerate}

\begin{figure}[h]
  \begin{center}
    \leavevmode
 \centerline{\epsfig{file=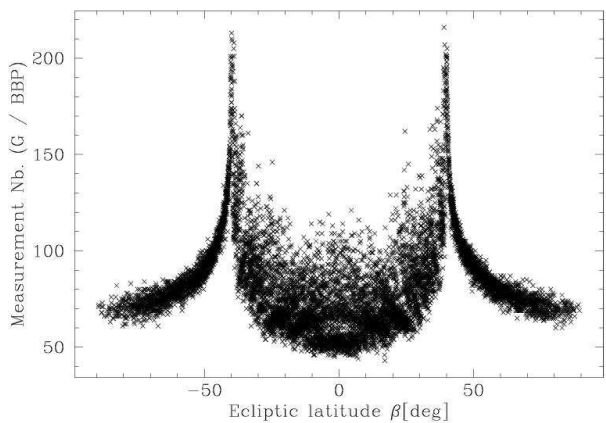,width=1.0\linewidth}}
   \end{center}
  \caption{Number of measurements (transits) as function of the ecliptic latitude $\beta$.
 The mean number of measurements is 82.}
 \label{fig:nmesbetaparis}
\end{figure}

The Astro-field will gather two observing directions separated by an angle of 106 degrees: the preceding (PFOV) and following (FFOV) fields of view. In the Astro-field, a star has measurements  generally grouped consecutively  with the time intervals of 1h46, 4h14, 1h46, etc, corresponding to observations in the PFOV-FFOV-PFOV, etc.
Then the next group will occur typically 30 days later. Both the total length of the sequence of
short consecutive measurements and the time intreval between the grouped measurements
depend on the position of the star on the celestial sphere.
We show in Figure~\ref{fig:nmesbetaparis} the dependence on the number of measurements as
function of the ecliptic latitude $\beta$. The mean number of measurements is 82 but varies
between 45 to 210.

\subsection{Number of variable objects detected by Gaia}

The number of variable objects detected by Gaia is still quite uncertain. First estimations
done by \citet{LEJC00} lead to about 18 million classical variable stars.

Caution is still needed: for instance the estimates of the number of eclipsing binaries
by different authors differ greatly: Soederhjelm (private communication) estimated this number
to be half a million, whereas \citet{LEJC00} estimated it to 2-3 millions and \citet{TZ02} to 7
millions.We list several estimates for other variable types:

-- \citet{LEJC00} estimated the numbers of RR Lyr to 70\,000, of Cepheids to 2\,000-8\,000, 
    of Mira-LPVs to 200\,000, of $\delta\,$Scuti stars to 60\,000-240\,000, of SPBs to 15\,000,
    of $\beta\,$ Cep stars to 3\,000.\\
-- Robichon (2002, private communication) estimated the number of planetary transit systems
   to 5\,000-30\,000.\\
-- \citet{VBWE03} estimated the number of Supernovae to 21\,000.\\
-- \citet{TTAP02} estimated that Gaia will observe 720 optical counter
       parts of gamma ray bursts.

\subsection{Rate of correct detection of Periods}

Eyer and Mignard has investigated the rate of correct detection of the period
for a given periodic signal observed with the Gaia sampling law.
Sinusoidal signals have been generated for pairs of $P$ (period) and $Q$ (amplitude
to noise ratio) and sampled with the Gaia scanning law for 10\,000 stars uniformly
distributed over the celestial sphere. For each star a periodogram is constructed,
and the highest power period $P_h$ is confronted to the true period $P$.
About 1.4 million periodograms have been computed.

As expected, the rate of correct period detection depends on the number of measurements
and therefore on the ecliptic latitude.  Once the number of measurements effect is corrected,
the dependence on the ecliptic latitude does not fully vanish. A dip reaching 10--15\%
in the period correct detection rate is still present at ecliptic latitude 20 degrees.

Very high amplitude to noise ratio $Q$ is not necessary to recover the period.
Already at $Q=1.3$, the rate of correct detection is mostly above 90\%. For example a
sinusoidal signal of amplitude 3 mmag at $V=14$, will have its period correctly recovered
from several hours to hundreds of days.

\subsection{Variable Star Working Group activities}

In the previous sections we covered some activities of the Working Group.
They are fully described at the following website:
\begin{verbatim}
http://obswww.unige.ch/~eyer/VSWG/
\end{verbatim}

You may find more information on:\\
{\it -- Variable star characteristics:} We prepare a uniform description of variable star types.
Information collected are: general references, physical properties, variability properties
(times scales and amplitudes of the photometric and radial velocity variations), physical
source of variability, number of star known, fraction of variable is the same area of the HR diagram.
Once the filters of Gaia will be defined, variability at the filter wavelengths will be
estimated for each type. Such characteristics will serve  for completing the Galactic Model
used for GDAAS (Gaia Data Access and Analysis Study),  for the estimation of the number
of variables, as well as for the classification.

{\it -- Compilations of past, present and future surveys:}
We want to keep track of the research done on these surveys. Therefore we give a list of past surveys with their returns on variable stars. We will add the methods used to analyse the data. We will also give the characteristics of some future surveys.

{\it -- Forecasts about variable stars to be detected by Gaia:}
We gather published information from different researchers on the forecasts.

{\it -- Tools:} We give a list of typical tools which are used in variable star research.
Algorithms of Period search, time scales, or classification methods. We put online software
for some algorithms, or links to the codes whenever possible.

{\it -- GDAAS related activities:}
We are developing software for the GDAAS analysis (see the following section).

\section{Global analysis}

\begin{figure}[h]
  \begin{center}
    \leavevmode
 \centerline{\epsfig{file=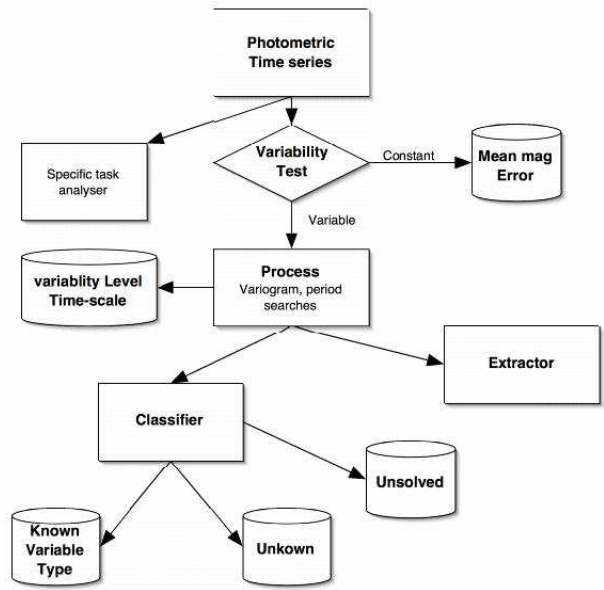,width=1.0\linewidth}}
   \end{center}
  \caption{Organigram of used data processing of the variable stars,
    baseline of the current development of the software for GDAAS.}
  \label{fig:organigram}
\end{figure}

The variability analysis can be seen as a process of data compression.
If a star has a constant magnitude, then the whole time series can be summarised by
two quantities (per band):
      the mean magnitude and its precision.
For instance the compression factor for the G-band photometry with 80 measurements is about
140 (the measurement epochs can be removed). If the object is detected as variable, then
the task to compress the data is more laborious, the time series may be reduced, for example,
to amplitude(s), time-scale(s), period(s), epochs, parameters of Fourier series, variable types.
The goals of the global analysis are to separate the observed objects in constant (or more
precisely non detected as variable) and variable, and to reach a classification with the
associated relevant parameters. We proposed an organigram to fulfil these goals
in Figure~\ref{fig:organigram}.

\subsection{Detection of variability}

The first step is to detect variability. As different variability behaviours have different
photometric signatures, a set of different statistical tests had to be designed.

Thanks to the very high G-magnitude precision, it has been decided, during the VSWG
Cambridge meeting in 2004, to use the magnitudes deduced from one passage
of the observed object over the 11 CCDs of the Astro-field. Studies of very short term
variability should also be explored using the per CCD information (most algorithms developed
for the per transit analysis are applicable at this level). 

To detect variability we perform a number of statistical tests on the G-band photometric data.
For these tests two aspects are important:
\begin{enumerate}
 \item they should be mathematically formulated in a cumulative way, so that during the
    mission, when new data are coming, the values of the statistical tests can be
    updated from old cumulated quantities and the new data. So, the variability status of
    every object can be monitored.
 \item for each test and observed time series, p-values will be computed.
   In the framework of hypothesis testing under the null hypothesis, a p-value is the probably
   of observing a value more extreme than the one actually observed. It allows to set a 
   variability detection threshold independent of the number of measurements and of the noise
   level.
\end{enumerate}
   
The chosen tests are the $\chi^2$ test, tests on the asymmetry and kurtosis (these tests
are related to three moments of the distribution),  Abbe test \citep{JvN41}, a test to
detect outliers, and a test on the presence of a slope.

The stellar variability for certain variable types may remain undetected in a wide
band filter like the G-band because of compensation effects in the spectral interval covered by the G-band.
The MBP and BBP bands will also be used. One method is to compute the covariant matrix between the different bands $M_{k,l}= \sum_{i=1}^n (m_{k,i}-\overline{m_k}) (m_{l,i}- \overline{m_l})/(\sigma_{k,i} \sigma_{l,i})$, where $n$ is the number of measurements, the indices $k, l$ refer to the available photometric bands and $\overline{m_l}$ the mean $l$-magnitude.
The matrix diagonal terms are the values associated to the $\chi^2$.  If we diagonalise
the matrix then we will get a signature of the physical nature of the variability.
It seems an elegant way to combine the information of different filters and to enhance the variability detection capability. We will perform some tests on the five Sloan filters of
SDSS data and also on simulated data generated from synthetic spectra.

\subsection{Modelling the variability}

Once the variability is detected. Some further processing is require to characterise 
the variability.

For strictly periodic behaviours, periodograms can be computed and the period of highest "power" can be used. There is no method which is superior to all the others
for finding the period of a periodic signal, some are performing better for certain types of
behaviour, some others are more less prone to certain types of aliasing. The CPU consumption
time may be also very different, though some optimisation are possible.
We plan, as activities of the working group, to start benchmarking of these different methods.
Simple method to recognise mono-periodic and multi-periodic phenomena should be also planed.

Fourier series are adequate mathematical representation to model periodic  variability
(though often problematic for EA eclipsing binaries). However it appears difficult to
separate a set of well behaved light curves even after a Fourier decomposition with analysis
of residues \citep{LECB04}. 

If the star is irregular, we have to rely on more general description:

-- The amplitude of variation: Several estimations are found in the literature. It is
customary to give magnitudes at minimum and maximum light, however these quantities are not
robust to outliers, and are meaningless if the amplitude of variation is small compared to
the noise measurements. There are other estimations: signal variance, inter-quartile
(-decile) range, estimated amplitudes correct from noise measurements.

-- Time scales: The power spectrum can be used as a general description of the signal behaviour
   \citet{WEVB04}. Calculation of variogram or auto-correlation algorithms maybe useful tools
   to provide time scales, when both the variable object and the time sampling are irregular

These questions will be addressed.

\subsection{Classification}

We distinguish between two different strategies: a global approach to classification
and an extraction of certain variable types.

Mostly what has been done in the microlensing surveys is to extract some specific
variability types. Criteria have been designed for known variable types, some few examples:
RR Lyrae stars \citet{CA00}, R Coronae Borealis Stars \citet{CA01}, etc, from MACHO data, 
Cepheids \citep{AU99}, eclipsing binaries \citep{LW04}, etc, from OGLE data.
On Hipparcos data, multivariate discriminant analysis has been applied to extract Slowly Pulsating B stars, and $\gamma\,$Dor stars, cf. \citet{CWetal98}, \citet{CAetal98}.

For the moment very few attempts of global classification have been done, examples can
be found in
\citet{GP04} on ASAS data,
\citet{LECB04} on ASAS data,
\citet{VBetal03} on MACHO data,
\citet{PWetal01}, on ROTSE data,
Marquette (private communication) on EROS data.

\begin{figure}[h]
  \begin{center}
    \leavevmode
 \centerline{\epsfig{file=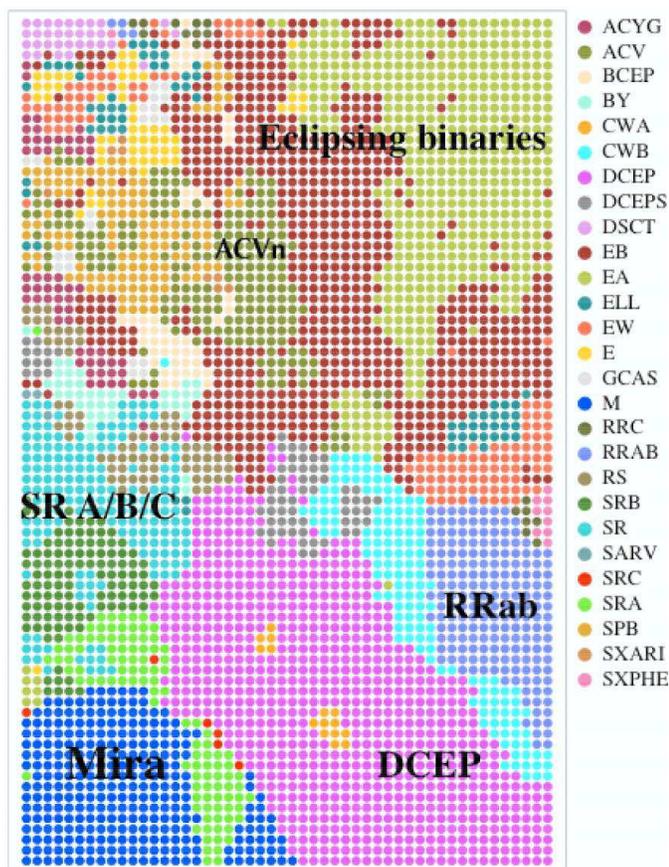,width=1.1\linewidth}}
   \end{center}
  \caption{Classification of {\sc hipparcos} periodic variable star catalogue with Self-Organising Maps by Dr A. Naud.}
\label{fig:soms}
\end{figure}

We are therefore nowadays in a stage of method exploration. Very promising methods like
Self-Organising Maps (SOMs) are studied, see \citet{WEVB04} for
a description of the method, cf. also \citet{DBetal04}. Dr A.~Naud (Torun Unversity) 
started a study on the global classification with SOMs of the Hipparcos periodic variable
star catalogue. This classification is using for each variable star its period, amplitude,
colour V-I and the skewness of the photometric time series. The  result is displayed in Figure~\ref{fig:soms}. We observe that the classes of Mira, SRs, Cepheids and eclipsing binaries form isolated groups. The smaller amplitude variables are somewhat mixed.

Several classification methods should be studied and compared: their efficiency, their
capability to separate classes, their ability to detect new classes, their error levels,
their robustness with respect to the sampling, their CPU consumption, and their level of
required human interaction. 

Once some objects are recognised, further analysis can be done. Global treatment of eclipsing binaries have been done on OGLE data by \citet{SWRW01}, \citet{SWRW02} and are investigated  for the Gaia mission by \citet{APTZ04}.

Many tests can be run on existing surveys like ASAS, MACHO, EROS, OGLE. However we remind that
the wealth of Gaia data is larger, the precision is higher and as for Hipparcos, the sampling
is peculiar. Global classification studies will have to be done with GDAAS.

The global classification is not excluding procedures of specific tasks, selecting unusual and
rare objects. As an example of an special selection procedure, we mention the detected
{\sc hipparcos} planetary transits of HD\,209458, cf. \citet{NRFA00} and \citet{SS99}.
It could have been singled out in Hipparcos before the discovery  of \citet{DCetal00}.
The HD\,209458 {\sc hipparcos} photometric time series was not classified as variable,
the eclipse depth is very shallow with respect to the measurement noise, however with
a well tailored statistical test ran on a subsample of Hipparcos sources,  HD\,209458
is one of the best candidate for having possible transits.

\section{Conclusion}

We have started to write algorithms for GDAAS to analyse the photometric time series. If the
Gaia launch date of 2011 seems far away, the tasks required to analyse the Gaia data for
variability studies are enormous. If a general picture of the variability analysis
is present, many problems are not solved yet. In order to gauge correctly the efficiency of
proposed methods, of the analysis as a whole, there is the need to study thoroughly very
technical details.

There are also broader questions: how and when to deliver intermediate results on variable
objects during the mission, how to plan Earth follow-up observations, how to combine Gaia data
with coeval surveys, what tools to use for the visualisation of Gaia data and the quantities
results of the variability analysis.

If Gaia data will be an unprecedented data set, the variability analysis also requires
unprecedented ingenious methods, software developments and collaborative efforts of many
researchers.


%
%

\section*{Acknowledgments}

I would like to express thanks to many members of the Variable Star Working Group, who are
actively participating to the tasks defined within the working group. Particular thanks go to
Dr. Dafydd Evans (co-taskleader) and Dr. Francesca Figueras, who are of continuous help and support.

I would like also to thank for helpful comments and discussions Prof. M.~Grenon,
Prof. F.~Mignard, Prof. D.~Koch and Dr. F.~Pont.

\bibliographystyle{aa}
\bibliography{eyer}

\end{document}